\documentclass[letterpaper]{article} % DO NOT CHANGE THIS
\usepackage[preprint]{aaai2027}  % DO NOT CHANGE THIS

\usepackage[colorinlistoftodos,prependcaption,textsize=small, disable]{todonotes}%,textwidth=1.5cm
\usepackage[hyphens]{url}  % DO NOT CHANGE THIS
\usepackage{graphicx} % DO NOT CHANGE THIS
\usepackage{natbib}  % DO NOT CHANGE THIS AND DO NOT ADD ANY OPTIONS TO IT
\usepackage{caption} % DO NOT CHANGE THIS AND DO NOT ADD ANY OPTIONS TO IT
\usepackage{algorithm}
\usepackage{algorithmic}
\usepackage{amsmath} 
\usepackage{amssymb}
\usepackage{newfloat}
\usepackage{listings}
\DeclareCaptionStyle{ruled}{labelfont=normalfont,labelsep=colon,strut=off} % DO NOT CHANGE THIS
\floatstyle{ruled}
\newfloat{listing}{tb}{lst}{}
\floatname{listing}{Listing}

\usepackage{booktabs}
\usepackage{multirow}

\title{Toward Collective-Centric Evaluation of Preference Inference for \\ Participatory Democracy}
\author {
    Pierre-Antoine Lequeu\textsuperscript{\rm 1}\equalcontrib, %\corresponding,
    Salim Hafid\textsuperscript{\rm 2}\equalcontrib,
    Paul Lerner\textsuperscript{\rm 1}, Nazanin Shafiabadi\textsuperscript{\rm 1}, Laurène Cave\textsuperscript{\rm 3}, \\ David Mas\textsuperscript{\rm 4}, Jean-Philippe Cointet\textsuperscript{\rm 2}, Benjamin Piwowarski\textsuperscript{\rm 1}, François Yvon\textsuperscript{\rm 1} 
}
\affiliations {
    \textsuperscript{\rm 1}Sorbonne Université, CNRS, ISIR, Paris, France\\
    \textsuperscript{\rm 2}Sciences Po, médialab, Paris, France\\
    \textsuperscript{\rm 3}Sorbonne Université, STIH/CERES, Paris, France\\
    \textsuperscript{\rm 4}Make.org\\
    lequeu@isir.upmc.fr, salim.hafid@sciencespo.fr
}

\begin{document}

\maketitle

\begin{abstract}
To scale up collective decision-making, participatory democracy platforms such as Polis and Remesh enable online deliberation among thousands of participants. However, at this scale, participants cannot review every opinion submitted by others, producing highly sparse voting data that misrepresent patterns of consensus, conflict, and minority support. Platforms therefore increasingly rely on \emph{Preference Inference (PI)} models to predict missing votes. Yet this automation is not neutral: inferred preferences can artificially amplify, suppress, or reorder existing patterns of support, ultimately reshaping how the outcomes of a deliberation are interpreted. More generally, we lack a systematic understanding of how existing \emph{PI} methods affect the \emph{collective} preference landscape. To address this gap, we benchmark several existing \emph{PI} approaches in this context. Moving beyond conventional \emph{user-centric} evaluations centered on the accuracy of individual predictions, we introduce a \emph{collective-centric} evaluation framework that measures whether inferred votes preserve salient properties of the broader preference landscape. We further contribute the largest multilingual dataset of its kind: four consultations spanning over 90k participants, 1M votes, and 22 languages. Our experiments show that models with comparable predictive accuracy can differ substantially in the degree to which they preserve the collective structure. These results demonstrate that accuracy alone is insufficient for evaluating \emph{PI} in democratic settings. By contributing a novel comprehensive and \emph{collective-centric} evaluation benchmark for the task of \emph{PI}, this work aims to support the development of AI systems that scale deliberation without compromising the integrity of its democratic outcomes.
%\footnote{datasets available at \url{https://huggingface.co/collections/democratic-commons/citizens-consultations}}
\end{abstract}

\begin{links}
    \link{Code}{https://github.com/LequeuISIR/PrefInferenceEval}
\end{links}

%7 pages of technical content plus additional pages solely for references and the reproducibility checklist

\section{Introduction}

% expliquer ce qu'est la démocratie participative, une consultation citoyenne, pourquoi l'IA peut la rendre possible ou passer à l'échelle, quels sont les dangers
\begin{figure}
    \centering
    \includegraphics[width=\linewidth]{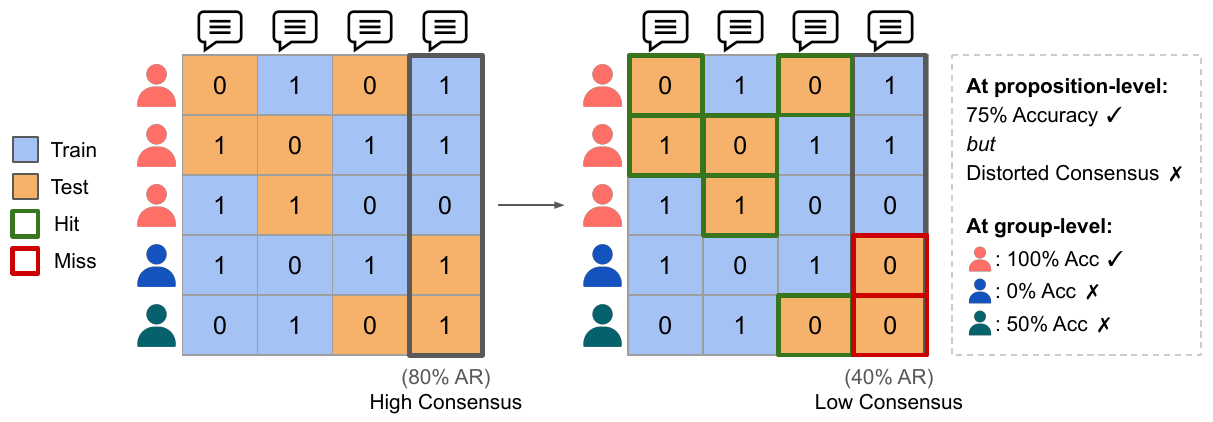}
    % \caption{\textbf{Accuracy fails to reveal distortions in the preference landscape introduced by \emph{preference inference} in online consultations.} In this example, Preference Inference correctly predicts 75\% of held-out test votes. In deployment, these test votes would correspond to unobserved entries that the model must infer. Despite this accuracy, inference reduces the approval rate (AR) of a highly consensual proposition from 80\% to 40\% and distributes errors unevenly across sociodemographic groups. This motivates evaluating \emph{proposition-level} support patterns and \emph{participant-level} group outcomes alongside individual prediction accuracy.}
    % More concise:
    \caption{\textbf{Accuracy fails to reveal distortions induced by \emph{Preference Inference}.} In this example,  75\% of held-out test votes (unobserved in deployment) are correctly inferred. Despite this accuracy, inference drops a consensual proposition's approval rate (AR) from 80\% to 40\% and distributes errors unevenly across socio-demographic groups. This motivates evaluating \emph{proposition-level} support and \emph{participant-level} outcomes alongside accuracy.}
    \label{fig:summary}
\end{figure}

Participatory
%JPC: Je parlerais plutôt de participatory democracy car beaucoup de ces plateformes ne sont pas délibératives
democracy grounds the legitimacy of collective decisions in the capacity of citizens to participate in forming those decisions \cite{habermas2015between, rawls1997idea, dryzek2003social}. Participatory democracy platforms such as Polis and Remesh %, and Make.org %c'est okay d'avoir make dans la liste?
promise to extend such participation beyond small, resource-intensive forums to large-scale consultations used in public policy design, participatory budgeting, and peace-building \cite{konya2025using, fishkin2019deliberative, landemore2020open, small2021polis}. In these consultations, participants submit propositions in free form text and express support for or opposition to propositions submitted by others, resulting in a collection of textual propositions and a participant–proposition voting matrix (see Figure~\ref{fig:summary}).%: in practice, the ``test'' votes must be inferred). 

However, given the volume of inputs, each participant can only engage with a small fraction of all propositions, leading to a highly sparse voting matrix. This sparsity creates the need for \emph{Preference Inference (PI)}: the task of predicting whether a participant would support or oppose a proposition they have not reviewed \cite{konya-etal-2022-elicitation}. These predictions carry concrete implications as they reconstruct the \emph{landscape of preferences} subsequently used to identify opinion groups, select representative propositions, and produce summaries for facilitators and decision-makers \cite{small2021polis,revel2025representative,fish2026generative}. Prior work on deliberation summarization demonstrates that different models can produce substantially different levels of proportional representation and may erase minority voices \cite{zhu2025can,hafid2026algorithmic}. Distortions introduced during \emph{PI} may thus propagate throughout the deliberative chain and alter the resulting collective representation.% reports that decision-makers ultimately receive.

% This raises a normative challenge. Agonistic and pluralist theories of democracy emphasize that disagreement is an essential and productive feature of democratic politics \cite{crawford2016can, mouffe1999deliberative}. 
Since \emph{PI} errors may be distributed unevenly (for instance, as a function of support size or internal group-level correlations), certain propositions risk being poorly modeled, generating spuriously consensual or non-consensual outcomes \cite{revel2026ai}, or under-representing specific socio-demographic and interest groups.
\emph{PI} should therefore neither erase genuine political conflict to produce an appearance of agreement nor create an illusion of disagreement by fragmenting groups or distorting propositions around which participants genuinely agree. Deliberative theory offers a useful middle ground through the notion of \emph{meta-consensus}: deliberation need not eliminate disagreement, but should accurately structure the values, judgments, and preferences over which disagreement persists \cite{dryzek2012foundations}. A democratic \emph{PI} model must therefore operate between two failures: artificial consensus and artificial conflict \cite{revel2026ai}. We argue that it can achieve this by preserving the salient structure of the expressed \emph{landscape of preferences} at two levels: how \emph{propositions} are positioned through patterns of support and opposition, and how \emph{participants} are organized into behavioral and socio-demographic groups.

% This latter risk raises an equality concern that certain groups’ preferences or that certain propositions are inferred with less accuracy than others'. Moreover, using algorithms to infer individual preferences also creates an accountability problem for the entire participatory process, whose legitimacy is lowered when preferences are not directly resulting from participation but this problem is out of the scope of our article. 
Meeting this dual objective creates a corresponding evaluation problem: how can we determine whether an inference model preserves, rather than distorts, the proposition- and participant-level structures of the preference landscape? A preference matrix is expected to exhibit a clustered structure because political preferences are often organized by relatively stable shared belief systems  \cite{converse2006nature}. Yet, \emph{PI} systems are commonly assessed using classification or recommendation metrics that aggregate errors across participant–proposition pairs \cite{konya-etal-2022-elicitation}, overlooking whether inference alters proposition-level support patterns, group-specific approval profiles, or vote-based community structure. More broadly, research on fairness in inference systems has emphasized the need to connect abstract metrics to the normative requirements of particular socio-technical contexts \cite{deldjoo2024fairness}. As illustrated in Figure \ref{fig:summary}, a model may achieve high accuracy while still distorting the preferences of several socio-demographic groups. We therefore argue for complementing standard \emph{user-centric} metrics with \emph{collective-centric} metrics designed to assess whether inference preserves the landscape of preferences.\footnote{Algorithmic preference inference also introduces an accountability challenge when preferences do not stem directly from participation, though addressing this limitation lies beyond our scope.}

To address this gap, this work makes three contributions:

\noindent \textbf{(1) }We construct a theoretical and empirical framework for evaluating how Preference Inference reshapes the \emph{landscape of preferences} derived from a voting matrix. We formalize this landscape along two complementary dimensions grounded in social sciences: the positioning of \emph{propositions} through patterns of support and opposition, and the grouping of \emph{participants} through voting behavior and socio-demographic attributes. Building on this formalization, we develop an evaluation benchmark that complements standard predictive metrics by measuring whether inferred preferences preserve proposition-level approval rates (AR) and participant-level preference structures, including vote-based communities and socio-demographic group patterns.

\noindent \textbf{(2)} We contribute a new multilingual dataset comprising four  consultations obtained via a partner platform. To our knowledge, it is the largest corpus of its kind, containing a total of almost 90k participants, 1M votes, and 22 distinct languages. To support reproducibility and facilitate the use of our metrics, we make our code and dataset publicly available.%\footnote{See the ZIP in Supplementary Material}

\noindent \textbf{(3)} We conduct an extensive evaluation of a comprehensive set of Preference Inference models, showing that models with high predictive accuracy can still substantially distort patterns of consensus and preferences of particular socio-demographic groups.

\section{Related Work}\label{sec:rel-work}
% \todo[inline]{lier au stance detection? sorte de multi-target stance detection mais où les targets sont dynamiques?} 

%\subsection{Preference Inference in Participatory Democracy Platforms} 
\subsection{Participatory Democracy Platforms} 
Preference Inference operates within the broader pipeline through which participatory democracy platforms organize, represent, and summarize large-scale public input. These platforms typically represent participants’ expressed preferences as a participant-proposition voting matrix, and use different inference strategies to deal with sparsity. Polis imputes missing entries using proposition-level mean votes before applying dimensionality reduction and clustering \cite{small2021polis}, whereas Remesh combines language model representations with latent factors learned from observed voting behavior \cite{konya-etal-2022-elicitation,konya2023deliberative}. The resulting inferred matrices support downstream tasks such as opinion group identification or automated summarization. Such systems have seen various high-stakes deployments: Remesh by the UN for peace-building \cite{ovadya2023generative}, and Polis in Taiwan to shape Uber and taxi regulations \cite{konya2023deliberative}.

From a social-science perspective, Preference Inference goes beyond a mere technical operation: it attributes preferences that participants have not explicitly expressed, thereby shaping how citizens and groups become represented within the consultation and how collective preferences are made visible to decision-makers \cite{revel-etal-2025-aifacilitated}. PI should therefore be understood as both a predictive mechanism and a form of algorithmic representation. 
%Drawing on Innerarity’s framework for evaluating predictive analytics in terms of equality, accountability, and legitimacy \cite{innerarity2023predicting}, we focus particularly on equality, asking whether some groups’ preferences are inferred less accurately than others. 
Inspired by Innerarity’s discussion of the normative implications of predictive analytics \cite{innerarity2023predicting}, we focus on one aspect of algorithmic fairness: whether prediction accuracy differs systematically across socio-demographic groups and vote-based communities. Building on this perspective, we introduce novel metrics to assess whether inferred votes preserve proposition-level support patterns and whether preferences are inferred with comparable accuracy across groups.
%Building on this perspective, we operationalize these normative concerns through systematic evaluation, introducing novel metrics that assess whether inferred votes preserve proposition-level support patterns and whether socio-demographic groups’ preferences are inferred equally accurately. 
 Moreover, because multilingual data remain underexplored in existing PI benchmarks, we evaluate our framework on a new multilingual dataset comprising four consultations, over 90K participants, 1M votes, and 22 languages.

%Building on this perspective, our work operationalizes these normative concerns into systematic evaluation by introducing metrics that assess whether inferred votes preserve proposition-level support patterns and participant-level behavioral and socio-demographic structures.

% structure
% \begin{itemize}
%     \item overview of how these platforms typically work
%     \item overview of societal relevance of these platforms (Polis' example with Uber)
%     \item Why do we even need to infer citizens' preferences? Overview of how platforms work and why inference is required for upscaling online consultations/deliberations
%     \item Overview of how inference preference affects a series of other downstream tasks
%     \item Overview of the social science perspective on inferring votes (Innerarity, 2023), (Crawford, 2016), (Revel et Pénigaud, 2025)
% \end{itemize}

\subsection{%Evaluating 
Preference Models for Collective Representation}

\paragraph{Models}
Although Preference Inference is a cornerstone task for participatory democracy platforms, it long predates these systems and has a substantial history across several research traditions. Early probabilistic models from psychometrics and conjoint analysis represented preferences through latent utilities, pairwise comparisons, and the attributes of alternatives \cite{thurstone1927law, green1978conjoint}. Later work introduced adaptive survey methods that collect informative responses without requiring every participant to evaluate every proposition \cite{salganik2015wiki}. More contemporary approaches draw on recommender systems and related latent-variable methods for user-item data, including matrix-factorization and collaborative filtering \cite{koren2009matrix}, latent-factor models \cite{johnson2014logistic}, and regularized matrix completion \cite{bilich2023faster}. These methods estimate missing responses from shared low-dimensional structure across participants and items. 
Recent enhancements additionally incorporate the semantic content of propositions through language model representations \cite{konya-etal-2022-elicitation}, following recent advances in NLP \cite{sorensen2025value} and stance detection \citep{allaway-etal-2020-zeroshot,barriere2023multilingual}. 
\citet{blair-etal-2026-embeddings} go further by training language models so that distances accurately represent participants' preferences.
Across methods, Preference Inference must typically contend with sparsity \cite{koren2009matrix} and cold-start participants and propositions \cite{lee2019melu}. In this work, we show that multilingual content poses an additional challenge.

\paragraph{Evaluation}
% In recommender systems, Preference Inference models are generally evaluated using classification or ranking metrics aggregated across user-item pairs. Evaluation has expanded beyond accuracy to consider user- and item-side fairness in order to reveal whether predictive accuracy or utility differs across predefined user groups \cite{deldjoo2024fairness, zhao2025fairness, ge2010beyond}. 
In recommender systems, Preference Inference models are generally evaluated using classification or ranking metrics aggregated across user-item pairs, with recent work expanding to user- and item-side fairness \cite{deldjoo2024fairness, zhao2025fairness, ge2010beyond}.
%\newline
However, participatory democracy demands more than the equal predictive performance common in recommender systems, as group-level parity does not guarantee faithful collective representation. For example, suppose Group A supports proposition 1 and opposes proposition 2, while Group B holds the opposite preferences. A model that predicts 50\% support for both propositions in both groups will show no disparity when evaluated using a conventional group-fairness metric, because both groups are inferred equally inaccurately. Yet the inference has erased the central \emph{structure} of the consultation: the groups originally held opposite preferences, but now appear identical. To address this gap, we introduce metrics that evaluate whether inference preserves such group-specific approval profiles and vote-based community structure. Put simply, in addition to assessing whether groups are predicted equally well, our metrics determine whether the inferred voting matrix still represents what the groups believe and how they differ.

\section{Theoretical Framework}
% \todo[inline]{Formalisation version originale}

% \paragraph{Notations} 
% Let us consider a set of $N$ participants (users) $\mathcal{U} = \{u_i\}_{i \leq N}$ and a set of $M$ propositions (thoughts) $\mathcal{T} = \{t_i\}_{i \leq M}$ on a democratic process $\mathcal{D}$. The opinion on user $u_i$ on proposition $t_j$ is denoted $v_{ij}$. We define $\mathcal{V}_{true} = \mathcal{U} \times \mathcal{T} \times \{-1,1\}$\footnote{For simplicity, we also use the $\mathcal V$ notation for the $N \times M$ voting matrix.}\todo{Plus simple de ne l'appeler que V en notant qu'une partie est observée et une partie non observée -> (Salim) j'ai reformulé plus bas en allant dans cette direction} as the set of opinions of all participants on all propositions, with $v_{ij}=+1$ (resp. $-1$) when $u_i$ agrees (resp.\ disagrees) with $t_j$. The democratic process provides a subset $\mathcal{V}_{known} \subset \mathcal{V}_{true}$ of opinions expressed by the participants during $\mathcal D$. Elicitation Inference aims at predicting the set of unknown opinions $\mathcal{V}_{unk} = \mathcal{V}_{true} \setminus \mathcal{V}_{known}$. We denote $\mathcal{M}$ the EI model predicting $\mathcal{V}_{unknown}$ and $\mathcal{M}(\mathcal{V}_{unk})$ the predicted set of votes. Finally, we define $\mathcal{V}_{infer}^\mathcal{M}=\mathcal{V}_{known} \cup \mathcal{M}(\mathcal{V}_{unk})$ as the "filled" voting matrix using $\mathcal{M}$.

% \todo[inline]{Formalisation version Salim}
\paragraph{Notation}
Consider a democratic process $\mathcal{D}$ involving a set of $N$ participants
$\mathcal{U}=\{u_i\}_{i=1}^{N}$ and a set of $M$ propositions
$\mathcal{T}=\{t_j\}_{j=1}^{M}$. Participants' opinions are represented by a
voting matrix $\mathbf{V}\in\{-1,+1\}^{N\times M}$, where $v_{ij}=+1$ if
participant $u_i$ supports proposition $t_j$, and $v_{ij}=-1$ if they oppose it.
Because participants vote on only a subset of propositions, the democratic
process provides a partially observed matrix
$\mathbf{V}_{\mathrm{obs}}\in\{-1,+1,\bot\}^{N\times M}$, where
$v^{\mathrm{obs}}_{ij}=v_{ij}$ when the vote is observed and
$v^{\mathrm{obs}}_{ij}=\bot$ otherwise.

\paragraph{Problem formalization: Preference Inference}
Given the partially observed voting matrix $\mathbf{V}_{\mathrm{obs}}$, a
Preference Inference model $\mathcal{M}$ predicts the missing entries and
produces a completed voting matrix
$\hat{\mathbf{V}}^{\mathcal{M}}\in\{-1,+1\}^{N\times M}$, defined by

% \[
% \hat v_{ij}^{\mathcal{M}}=
% \begin{cases}
% v^{\mathrm{obs}}_{ij},
% & \text{if } v^{\mathrm{obs}}_{ij}\neq\bot,\\[4pt]
% \mathcal{M}(\mathbf{V}_{\mathrm{obs}},u_i,t_j),
% & \text{if } v^{\mathrm{obs}}_{ij}=\bot.
% \end{cases}
% \]

\begin{equation}
\hat v_{ij}^{\mathcal{M}} =
\left\{
\begin{array}{ll}
v^{\mathrm{obs}}_{ij}, 
& \text{if } v^{\mathrm{obs}}_{ij}\neq\bot, \\[4pt]
\mathcal{M}(\mathbf{V}_{\mathrm{obs}},u_i,t_j), 
& \text{if } v^{\mathrm{obs}}_{ij}=\bot.
\end{array}
\right.
\end{equation}

Preference inference therefore aims to estimate the unobserved votes while
leaving the observed votes unchanged. In this work, we evaluate not only the
accuracy of these predictions but also whether
$\hat{\mathbf{V}}^{\mathcal{M}}$ preserves the collective properties of the
underlying voting matrix $\mathbf{V}$.

%\todo[inline]{Version Salim (landscape of preferences)}
\subsection{Characterizing the Landscape of Preferences}\label{ssec:collective}
We define the \textbf{landscape of preferences} as the set of collective structures derived from the voting matrix. We characterize it through two complementary perspectives:
% To evaluate the effect of the inferred voting matrix on the \emph{landscape of preferences}, we must first define what this landscape is. We do so by characterizing the main structures induced by participants' votes. We consider two complementary perspectives: 
how \textbf{propositions} are positioned relative to participants' preferences, and how \textbf{participants} are organized through their voting behavior and social attributes.

\paragraph{Propositions}
On the proposition side, the voting matrix induces aggregate patterns of support and opposition. We define the \emph{Approval Rate} of a proposition $t_j$, denoted $\mathrm{AR}(t_j)\in[0,1]$, as the proportion of positive votes it receives. Approval rates position propositions along the landscape of preferences: some propositions are broadly supported, some are broadly rejected, and others are contested, with approval rates close to $0.5$. These proposition-level patterns are central to how consultations are interpreted, since they indicate which ideas attract consensus, opposition, or disagreement. A Preference Inference model should therefore not artificially inflate, suppress, or reorder these patterns.

\paragraph{Participants}
On the participant side, the voting matrix induces both behavioral and socio-demographic structures. First, participants may form vote-based \textit{communities}\footnote{We borrow the \textit{community} terminology from graph theory, as further explained in the next section.} \cite{mason-2022-uncivil}: groups of users with similar voting profiles, characterized by high within-group similarity and lower between-group similarity. These communities capture distinct perspectives within the consultation and reflect the plurality of preferences expressed by participants. Second, participants may belong to socio-demographic groups defined by attributes such as age, gender, ethnicity, or location. Because inference may affect these groups differently, Preference Inference should be evaluated with respect to whether it preserves their position and representation in the landscape of preferences, rather than amplifying, attenuating, or erasing group-specific patterns.

\subsection{Measuring the Distortion of Preferences}
\label{subsec:metrics}
We select metrics that evaluate whether preference inference preserves the landscape of preferences along its two main dimensions: propositions and participants. On the proposition side, we use \emph{Balanced Accuracy} to measure whether individual missing votes are predicted correctly despite label imbalance, and \emph{Approval Rate Hit Ratio} to assess whether the resulting patterns of support and opposition across propositions remain statistically credible. On the participant side, we measure \emph{Consensus Distortion} to evaluate whether inference alters the approval patterns of particular participant groups, and \emph{Community Stability} to assess whether groups of participants with similar voting behavior remain represented after matrix completion. We complement these measures with fairness metrics for socio-demographic groups: \emph{minimum group Balanced Accuracy} to capture the performance experienced by the worst-served group, and \emph{Equalized Odds Ratio} to measure disparities in error rates across groups. Together, these metrics provide a comprehensive assessment of whether inference preserves both how propositions are positioned and how participants (including behavioral communities and socio-demographic groups) are represented in the preference landscape. We define each metric below. 

\paragraph{Approval Rate Hit Ratio (ARHR)}
ARHR measures whether a PI model $\mathcal{M}$ provides statistically credible approval rates \textit{at the proposition-level}. For each proposition $t_j$, the observed approval rate $\mathrm{AR}(t_j|\mathbf{V}_{\mathrm{obs}})$ is treated as an estimate of the approval rate in the underlying full matrix $\mathbf{V}$. Considering votes on $t_j$ as Bernoulli trials (where success is seen as supporting a proposition), we compute a $95\%$ Wilson confidence interval\footnote{We use the Wilson score interval as it correctly handles cases with small sample size and extreme probability values. More details are given in the Supplementary Material.} \cite{wilson-1927-probable} from the observed votes in $\mathbf{V}_{\mathrm{obs}}$, denoted $\mathrm{CI}_{0.95}(t_j|\mathbf{V}_{\mathrm{obs}})$. The Approval Rate Hit Ratio is then, for a given PI model $\mathcal{M}$:
%\mathrm{ARHR}(\mathcal{M})
%=
\begin{equation}
\frac{1}{|\mathcal{T}|}
\sum_{t_j\in \mathcal{T}}
\mathbf{1}
\left[
\mathrm{AR}(t_j\mid \hat{\mathbf{V}}^{\mathcal{M}})
\in
\mathrm{CI}_{0.95}(t_j\mid \mathbf{V}_{\mathrm{obs}})
\right]
\end{equation}

A system with a low ARHR outputs statistically unlikely approval rates, meaning it is likely to distort the actual approval from the participant.
% \begin{equation}
% \mathrm{ARHR}(\mathcal{M})
% =
% \frac{1}{|\mathcal{T}|}
% \sum_{t_j\in \mathcal{T}}
% \delta_{
% \mathrm{AR}(t_j\mid \hat{\mathbf{V}}^{\mathcal{M}})
% \in
% \mathrm{CI}_{0.95}(t_j\mid \mathbf{V}_{\mathrm{obs}})
% }
% \end{equation}

% The observed AR of a proposition $AR(t_i|\mathcal{V}_{known})$ is an estimate of the true AR $AR(t_i|\mathcal{V}_{true})$.
% We can consider the voting task as a binomial trial experiment (with success=support and failure=oppose) and compute a confidence interval at 95\% of the true AR from the observed one. The AR Hit Ratio computes the proportion of propositions for which $AR(t_i|\mathcal{V}_{infer}^\mathcal{M})$ is in this interval. An AR Hit Ratio $\geq0.95$ is expected for statically credible results.\todo[author=FY]{Why is this here? I find this confusing to have two distinct refs. to 0.95 in the same para.}

\paragraph{Consensus Distortion (CD)} We conceptualize consensus on a proposition as the distribution of approval across relevant participant groups, following work that evaluates common ground through cross-group agreement \cite{small2021polis, konya2025using, hafid2026algorithmic}. Under this view, a proposition is seen as consensual only if its support is represented accurately across groups: substantially overestimating or underestimating the approval of any group may create a misleading picture of shared support or disagreement. CD therefore measures how much Preference Inference alters the \emph{group-level approval} profile on which judgments of consensus are based. %\todo{on pourrait aussi avoir un interval de confiance sur le consensus pre-inference} 
% CD measures whether the votes inferred by an PI model $\mathcal{M}$ make propositions appear more consensual or less consensual than they truly are. \todo{je pense cette phrase à changer, pour expliquer pourquoi c'est important d'évaluer le groupe le plus négatif sur la question}
Let $\mathcal{C}$ denote a partition of $\mathcal{U}$ where each user $u_i$ is assigned to a cluster $c_k \in \mathcal{C}$. $\mathcal{C}$ can be defined based on protected attributes, such as age or sex, or be inferred from the observed voting data. We denote $\mathrm{AR}(c_k,t_j|\mathbf{V})$ as the approval rate of the user-group $c_k$ on $t_j$. We define the group-level approval-rate distortion as: 

\begin{equation}
\Delta(c_k,t_j,\mathcal{M})
=
\left|
\mathrm{AR}(t_j \mid c_k, \hat{\mathbf{V}}^{\mathcal{M}})
-
\mathrm{AR}(t_j \mid c_k, \mathbf{V})
\right|
\end{equation}
Because a large distortion for even one group can produce a misleading impression of cross-group consensus, the proposition-level consensus distortion is defined as the maximum group-level distortion:
\begin{equation}
\mathrm{CD}(t_j, \mathcal{M})
=
\max_{c \in \mathcal{C}_j} \Delta(c,t_j,,\mathcal{M})
\end{equation}

% Finally, the democratic-process-level CD is the average of the proposition-level maxima over the set of propositions $\mathcal{T}'$:
% \begin{equation}
% %\[
% \mathrm{\mathrm{CD}}(\mathcal{D},\mathcal{M})
% =
% \frac{1}{|\mathcal{T}'|}
% \sum_{t_j \in \mathcal{T}'}
% \mathrm{CD}(t_j)
% %\]
% \end{equation}

Finally, the democratic-process-level consensus distortion $\mathrm{\mathrm{CD}}(\mathcal{D},\mathcal{M})$ is the average of the proposition-level maxima over the eligible set of propositions $\mathcal{T}'$.

Unlike Approval Rate Hit Ratio, which compares inferred approval rates to confidence intervals estimated from $\mathbf{V}_{\mathrm{obs}}$, Consensus Distortion is defined against the underlying matrix $\mathbf{V}$. In empirical evaluations, $\mathbf{V}$ is approximated by held-out votes. The two metrics also differ in their level of analysis: ARHR evaluates whether \textit{proposition-level} aggregate approval rates are preserved, whereas CD evaluates whether \textit{group-level} approval rates are preserved within each proposition before aggregating across propositions.

% \paragraph{$\mathcal V$ Rank Stability}
% The rank of the voting matrix $\mathrm{rank}(V_{true})$ can be interpreted as the number of \emph{voting patterns} or \emph{ideologies} in $\mathcal D$. If all propositions and participants are fully independent, then $\mathrm{rank}(\mathcal V_{true})=\min(N,M)$. Propositions and participants are however not independent: (1) multiple propositions can express similar ideas, and (2) some participants share similar voting profiles on multiple propositions, even if the latter give different ideas.\todo{Not clear} Thus, $\mathrm{rank}(\mathcal V_{true}) < \min(N,M)$. \newline 
% We contend that after inference, the rank of $\mathcal{V}^{\mathcal{M}}$ should not significantly differ from the rank of $\mathrm{rank}(V_{true})$: a larger rank would mean that $\mathcal{M}$ created new unobserved \emph{voting patterns}, while a lower rank \todo{est-ce que c'est possible un lower rank? R: bien sur, pourquoi pas?} would imply that $\mathcal{M}$ failed to discover existing patterns.

\begin{table*}[t]
    \centering
    \resizebox{\textwidth}{!}{
    \begin{tabular}{lllrrrlc}
    \toprule
\textbf{Platform} & \textbf{Dataset} & \textbf{Topic} & \textbf{\# Votes} & \textbf{\# Participants}  &  \textbf{\# Propositions}  & \textbf{\# Attributes} & \textbf{Lang} \\
\midrule
\multirow{7}{*}{\textbf{Polis}}
 & austria-climate & Energy & 133,048 & 1,756 & 1,039 & -- & DE \\                                                                                                   
 & bowling-green & BGKY Improvement & 226,136 & 2,031 & 896 & -- & EN \\                                                    
 & 15-per-hour-seattle & Seattle Minimum Wage Impact & 2,995 & 339 & 54 & -- & EN \\                                   
 & march-on & Operation Marching Orders & 536,923 & 6,289 & 2,160 & -- & EN \\                                                       
 & bank-farmland & Land Use and Conservation & 33,530 & 404 & 293 & -- & EN \\                                                                                       
 & vtaiwan.uberx & Taiwan Uber Regulation & 49,996 & 1,921 & 197 & -- & EN \\                                                     
 & bg2050 & Bowling Green Future & 14,341 & 126 & 371 & -- & EN \\

\midrule
\multirow{3}{*}{\textbf{Remesh}}
 & right-to-assemble & Right to Assemble & 7,547 & 307 & 1,296 & 4 & EN \\                                                                                                                 
 & foreign-intervention & US Foreign Relations & 4,520 & 308 & 767 & -- & EN \\

 & campus-protests & University Campus Protests & 6,066 & 309 & 1,020 & 7 & EN \\

\midrule
\multirow{4}{*}{\textbf{Make} (ours)}%Ours}
 & steuer-debate & Taxation & 16,204 & 1,050 & 25 & 4 & DE \\

 & gc-sante & Healthcare & 226,010 & 25,058 & 1,331 & -- & FR \\

 & ingerence & Foreign Digital Interference & 103,671 & 8,968 & 559 & -- & FR \\

 & eurhope & Future of Europe & 892,273 & 54,421 & 4,366 & -- & 22 \\

\bottomrule

    \end{tabular}}
    \caption{Datasets statistics. The following datasets comprise protected socio-demographic attributes as metadata: \textit{right-to-assemble}: age, ethnicity, gender, politics; \textit{campus-protests}: age, education, gender, income, living-area, politics, religion; \textit{steuer-debate}: age, gender, living-area, politics.}
    \label{tab:statistics}
\end{table*}

\paragraph{Community Stability (CS)} We previously defined \emph{communities} as groups of participants who display similar voting . Hypothesizing that such dynamics can be found in $\mathbf{V}_{\mathrm{obs}}$, these communities should not disappear after inference. Voting matrices can be viewed as \textit{Signed Bipartite Graphs} (SBG)—graphs with two distinct sets of nodes (participants and propositions) and edges existing only between the two sets (preferences). A \textit{Community} in graph theory refers to a cluster of nodes strongly interconnected and weakly connected to others. In the \emph{PI} setup, it corresponds to participants with similar voting behaviors. The \textit{Community Stability} metric evaluates how these communities evolve after inferring the unobserved preferences of their population. 
More specifically, we only consider the subset of users who expressed at least $N_{\min} = 10$ opinions on propositions, and apply agglomerative clustering\footnote{Agglomerative clustering is favored over other methods as it has no stochastic component.} \cite{mullner-2011-modern} on the signed bipartite spectral representation \cite{kunegis-etal-2010-spectral} of the sub-graph to find the optimal number of clusters $k^*$ maximizing the silhouette score \cite{rousseeuw-1987-silhouettes}. We then apply the same method with $k^*$ clusters on the same user subset, but on the fully inferred matrix $\hat{\mathbf{V}}^{\mathcal{M}}$. We compare the resulting clustering using the Adjusted Rand Index \cite{hubert1985comparing}. A higher score indicates community stability, while a low score hints at their disappearance.

\paragraph{Other Metrics}
We compute additional standard metrics. First, we use \emph{Balanced Accuracy} (\textbf{B-Acc})—the average of the True Positive Rate (aka recall) and True Negative Rate—as it is well-suited for imbalanced datasets \citep{5597285}. 
% \begin{equation}
%     \frac{\sum_i\sum_j \mathbf{1}[v_{ij}]\mathbf{1}[\hat v_{ij} v_{ij}]}{2\sum_i\sum_j \mathbf{1}[v_{ij}] } + 
%     \frac{\sum_i\sum_j \mathbf{-1}[v_{ij}]\mathbf{1}[\hat v_{ij} v_{ij}]}{2\sum_i\sum_j \mathbf{-1}[v_{ij}] }
% \end{equation}
%,where $\delta$ denotes the Kronecker delta.
%\todo[author=FY]{Define} 
%We favor it compared to accuracy (used by \citet{konya-etal-2022-elicitation}) as overall, the labels are strongly imbalanced, with on average (\% a check) of opinions being supported. 
% Additionally, we compute fairness-aware metric when having access to protected attributes: 
Moreover, when protected attributes are available, we compute fairness-aware metrics:
\textbf{Min. B-Acc} is simply the \emph{Minimum Balanced Accuracy} across group subsets \citep{Weerts_Fairlearn_Assessing_and_2023}, reflecting a Rawlsian principle of evaluating outcomes for the worst-served group \citep{rawls1971theory}.
% \begin{equation}
%     \min_{c \in \mathcal{C}}\mathrm{baccuracy}(c)
% \end{equation}
\emph{Equalized Odds Ratio} (\textbf{EOR}; \citeauthor{NIPS2016_6a9659fe}, \citeyear{NIPS2016_6a9659fe}) is defined as the minimum of the True Positive Rate and False Positive Rate ratios across groups, where each ratio measures the disparity between the largest and smallest rates. 
% The True (resp. False) Positive Rate ratio being defined as the ratio between the largest and smallest True (resp. False) Positive Rate across groups:
\begin{equation}
    \min\left(\frac{\min_{c \in \mathcal{C}}\mathrm{TPR}(c)}{\max_{c \in \mathcal{C}}\mathrm{TPR}(c)}, \frac{\min_{c \in \mathcal{C}}\mathrm{FPR}(c)}{\max_{c \in \mathcal{C}}\mathrm{FPR}(c)}\right)
\end{equation}
An EOR of 1 indicates that all groups achieve identical TP, TN, FP, and FN rates.
%\todo{TODO min-balanced accuracy and EOR}

\section{Experiments and Results}

\subsection{Datasets} 
We select two existing datasets from two distinct online citizen consultation platforms—namely, Polis\footnote{\url{https://github.com/compdemocracy/openData}} and Remesh.\footnote{\url{https://github.com/akonya/polarized-issues-data}}
% (see the Supplementary Material for details on the selection criteria)
To these, we add four newly compiled datasets sourced from Make\footnote{\url{https://huggingface.co/collections/democratic-commons/citizens-consultations}}, a large-scale international civic technology platform. We contribute this novel corpus alongside our work to address limitations in the existing datasets. First, it provides non-English evaluation data (French, German, and 22 languages in \textit{eurhope}\footnote{bg, cs, da, de, el, en, es, et, fi, fr, hr, hu, it, lt, lv, nl, pl, pt, ro, sk, sl, and sv (ISO 639-1)}), countering the English-heavy bias of existing resources and enabling future multilingual analyses. Second, it captures national and transnational policy debates—such as taxation and foreign digital interference—at a generally larger scale than previous localized datasets. Finally, it offers diverse structural properties for evaluation, ranging from highly dense consultation matrices (e.g., over 50\% interaction density in \textit{steuer-debate}) to massive, sparse environments (\textit{eurhope} alone comprises over 892k votes and 54k participants). Table~\ref{tab:statistics} reports the details on all used datasets.

To unify the datasets across the different platforms, we restrict our analysis to \textit{agree} and \textit{disagree} votes, discarding \textit{neutral} votes (unavailable in Remesh) and pairwise preferences (exclusive to Remesh). For protected socio-demographic attributes, to ensure that the computed statistics are reliable, we only consider groups with $|c|\geq25$. Statistics of the post-processed datasets can be found in the Supplementary Material. 
%\itodo{filtre métadonnées}

\subsection{Models}
We evaluate a comprehensive set of Preference Inference models spanning the main methodological families identified in the related work.

\paragraph{Collaborative Filtering (CF)} CF predicts preferences using historical interaction patterns rather than explicit metadata. User-based models (\textbf{CF-user}) infer unseen votes by identifying individuals with highly correlated voting histories. Proposition-based models (\textbf{CF-prop}) calculate similarity between the propositions themselves based on overlapping ratings. The algorithm then aggregates these proximity metrics across the interaction matrix to infer the preferences.   

\paragraph{Nearest Neighbor (NN)} NN uses the embeddings of propositions from pretrained encoders. It relies on the text embedding $\mathbf{e}$ of the propositions by assigning to $\hat v_{ij}$ the opinion $v_{ij*}$ of the closest proposition in the embedding space for which the user already expressed a preference:
\begin{equation}
    j* = \mathrm{argmax}_{j' \neq j} \mathbf{e}_{t_j}\cdot \mathbf{e}_{t_{j'}}\ \ \ \  \forall t \in \mathcal{T}_{u_i}
\end{equation}
where $\mathcal{T}_{u_i}$ is the set of propositions on which $u_i$ voted.

\begin{figure*}
    \centering
    \includegraphics[width=1\linewidth]{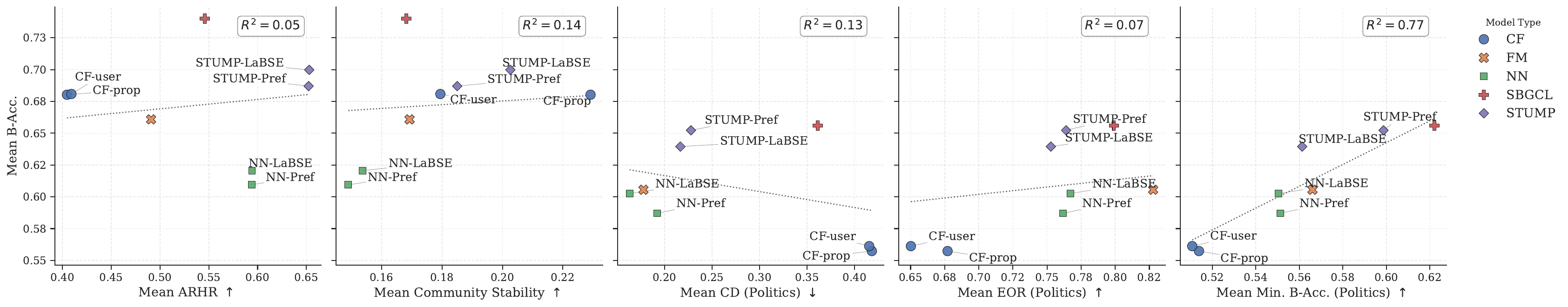}
    \caption{
    % Balanced Accuracy compared to ARHR, Community Stability, as well as CD, EOR and Min BA. for the \textit{politics} protected attribute. metrics are averaged per model across all datasets. For CD, EOR and Min. B-Acc, The average is done over the three datasets with protected attributes. Regressions are done over all (model, dataset) pairs, and not the displayed averages.
    B-Acc vs. ARHR, CS, CD, EOR and Min. B-Acc for the \textit{politics} protected attribute. CD, EOR and Min. B-Acc, are averaged over the three datasets with protected attributes and the other metrics are averaged over all datasets for each model. Regressions are computed over all (model, dataset) pairs, and not the displayed averages.
    }
    \label{fig:mean_scatter}
\end{figure*}

\paragraph{Factorization Machines (FMs)}
% FM learn representations of users and propositions from the existing preferences to infer new preferences. Standard factorization machines (\textbf{FM}) initializes both users and propositions randomly in $\mathbb{R}^P$ and learn these representation to optimize the utility of a pair $(u_i, t_j)=\sigma(u_i \cdot t_j)$  ($\sigma$ the sigmoid function) to be close to 1 if $u_i$ agrees on $t_j$, else to 0. 
FMs learn representations of users and propositions from existing preferences to infer new ones. Standard FMs initialize both user and proposition embeddings randomly in $\mathbb{R}^P$ and optimize them such that the utility for a pair $(u_i, t_j)$, mapped via the sigmoid function $\sigma(u_i \cdot t_j)$, approximates 1 if user $u_i$ agrees with proposition $t_j$, and 0 otherwise.

Additionally, we implement \textbf{STUMP} \cite{konya-etal-2022-elicitation}, an FM variant tailored for preference inference. STUMP initializes user  representations $u_i \in \mathbb{R}^P$ randomly\footnote{We experimented with initializing user representations from the embeddings of propositions they voted on, but this did not improve performance.} and uses pretrained proposition embeddings $\mathbf{e}_{t_j} \in \mathbb{R}^L$ as initialization. It then learns user embeddings alongside a linear projection matrix $\mathbf{W}_t \in \mathbb{R}^{L \times P}$ for propositions, and computes the utility of $(u_i, t_j)$ as $\hat v_{ij} =\sigma({u_i} \cdot \mathbf{W}_t^T \mathbf{e}_{t_j} )$, where $\sigma$ denotes the sigmoid function.\footnote{STUMP also learns a context-specific projection for consultations. As these contexts are only available on the Remesh datasets, we discard them and remove the projection.}

\paragraph{SBG link-prediction} As stated previously, consultations can be viewed as Signed Bipartite Graphs (SBG). Recent works have explored machine learning for link prediction in SBG. We implement \textbf{SBGCL} \cite{zhang-etal-2023-contrastiveb}, a  Graph Neural Network which uses a two-level view of the graph (the explicit inter-sets and implicit intra-set relation of nodes) and applies augmentation techniques and contrastive learning to learn representations of the nodes.
%\textbf{GegenNet} \cite{}\todo{ref} is a novel architecture using spectral decomposition and filtering and uses a convolutional network to learn node representations.

%\itodo{still mention stump+ and +- but briefly, say it did not improve}
% We also propose a modification of STUMP, namely \textbf{STUMP}$\pm$, which initializes user representations $e_{u_i} \in \mathbb{R}^M$ as the weighted sum of propositions on which the user voted, with a weight $+1$ for supported propositions and $-1$ for opposed propositions:
% We also propose \textbf{STUMP}$\pm$, a variant of STUMP that initializes user representations $\mathbf{e}_{u_i} \in \mathbb{R}^L$ as a weighted sum of propositions the user voted on, using weights of $+1$ for supported propositions and $-1$ for opposed ones:

% \begin{equation}
%     \mathbf{e}_{u_i} = \frac{1}{|\mathcal{T}_{u_i}|} \sum_{t_j \in \mathcal{T}_{u_i}} v_{ij} \mathbf{e}_{t_j}
% \end{equation}

% In STUMP$\pm$, the users' representations are not learned. Instead, we learn an additional projection $\mathbf{W}_u \in \mathbb{R}^{L \times P}$ alongside the already existing projection $\mathbf{W}_t$ and compute the utility as $\hat v_{ij} =\sigma(\mathbf{W}_u^T\mathbf{e}_{u_i} \cdot \mathbf{W}_t^T\mathbf{e}_{t_j} )$. 

\paragraph{LLM-based methods} Modern recommendation approaches leverage in-context learning and LLMs \cite{brown-etal-2020-language}. In this framework, an LLM is prompted with a simple instruction (cf. Supplementary Material) containing the user's observed positive and negative votes 
%it's ok everything fits \footnote{if it fits in the context window. Else, a subset is given} 
and asked to predict whether the user would approve of a new proposition (returning 1 for approval, 0 otherwise). The final prediction corresponds to the most likely token between the two\footnote{Because LLM recommendation is not scalable to large consultations (e.g. eurhope necessitates 250M inference), we only report its results on the three datasets with protected attributes.}:
\begin{equation}
    \hat v_{ij} = %-1+2\times  
    \mathrm{argmax}_{v\in\{0,1\}}\left(\hat P(v\ |\ t_j,\mathcal{T}_{u_i})\right)
\end{equation}

\subsection{Experimental Protocol}
For each consultation, votes ($\mathbf{V}_\mathrm{obs}$) are randomly partitioned into a 70-15-15 train-validation-test split. For learned models, we conduct an extensive grid-search over learning rates and model-specific parameters (see Supplementary Material for the search space and optimal configurations). We select the model checkpoint that maximizes \emph{B-Acc} on the validation set after a maximum of 400 training epochs, and evaluate its performance on the corresponding test set.
% For Balanced Accuracy, EOR, Min. Balanced Accuracy, and Consensus Distortion, we report results on the test set, a subset of $\mathbf{V}_\mathrm{obs}$. ARHR and Community Stability are computed on the full $\mathbf{V}^{\mathcal{M}}$ matrix.  
For B-Acc, EOR, Min. B-Acc, and CD, we report results on the test set, a subset of $\mathbf{V}_\mathrm{obs}$. ARHR and CS are computed on the full $\mathbf{V}^{\mathcal{M}}$.

We compare two text-embedding models: (i) \textbf{LaBSE} \citep{feng-etal-2022-language}, a 110M-parameter multilingual model; and (ii) \textbf{Pref} \cite{blair-etal-2026-embeddings}, a  1.2B-parameter model based on Sentence-T5 \cite{ni-etal-2022-sentencet5} that is specifically trained for preference representation.
For LLM recommendation, we use the open-weight post-trained model Qwen3-8B \citep{qwen3technicalreport}. % with a maximum of $k=5$ propositions for In-Context Learning examples, so as to fit the context size.

%We train each model on each consultation separately, using a 70-15-15 (?) train-dev-test random split. % on all the votes. 
%For trained models, we run a grid search over learning rates [...] and report the results for the best learning rate at the best checkpoint. \todo{a faire}  

\subsection{Results}
% All results are reported $\times 10^2$. We report per-model averages in this document. Detailed results are reported in the Supplementary Materials. We aim to answer two questions: (i) Do better performances in terms of balanced accuracy come with a less distorted landscape of preferences ? (ii) Do existing PI methods provide acceptable results regarding opinion distortion ?  
We present per-model averages throughout this section\footnote{All results are scaled by $\times 10^2$.}  (detailed breakdowns are available in the Supplementary Material). Our evaluation addresses two central questions: (i) Does higher balanced accuracy necessarily imply a less distorted landscape of preferences? (ii) Do existing PI methods preserve opinion structures adequately, or is further improvement necessary?
%\footnote{All results are scaled by $\times 10^2$.}

%\todo[inline]{ici, il faut qu'on clarifie pour chaque metrique, quelle $\mathcal{V}$ on utilise: le test, le inferred, etc...}
% \begin{table}[t]
%     \centering
  
%  \begin{tabular}{lrrr}
% \toprule
% Model & B-Acc. & ARHR & CS \\
% \midrule
% NN-Pref & 60.9 & 59.4 & 14.9 \\
% NN-LaBSE & 62.1 & 59.4 & 15.4 \\
% \midrule
% CF-user & 68.1 & 40.9 & 17.9 \\
% CF-prop & 68.0 & 40.5 & \textbf{22.9} \\
% \midrule
% FM & 66.1 & 49.1 & 16.9 \\
% STUMP-Pref & 68.7 & \underline{65.2} & 18.5 \\
% STUMP-LaBSE & \underline{70.0} & \textbf{65.3} & \underline{20.3} \\
% \midrule
% SBGCL & \textbf{74.0} & 54.6 & 16.8 \\

% \midrule
% % LLM & 60.3 & \textcolor{gray}{76.0} & \textcolor{gray}{8.5} \\
% %LLM & 61.0 & \textcolor{gray}{75.2} & \textcolor{gray}{5.7} \\
% \bottomrule
% \end{tabular}
%     \caption{Balanced Accuracy (B-Acc.), Approval Rate Hit Ratio (ARHR), and Community Stability (CS), averaged over all datasets. As the LLM does not scale to large-scale consultations, its ARHR and CS are computed only on a subset of 100,000 votes instead of the full voting matrix and are, therefore, not directly comparable. \itodo{ne pas oublier LLM}
% }% \todo[inline]{explain why LLM is in gray}}
%     \label{tab:mean_metrics}
% \end{table}

\begin{figure}[t]
    \centering
    \includegraphics[width=.9\linewidth]{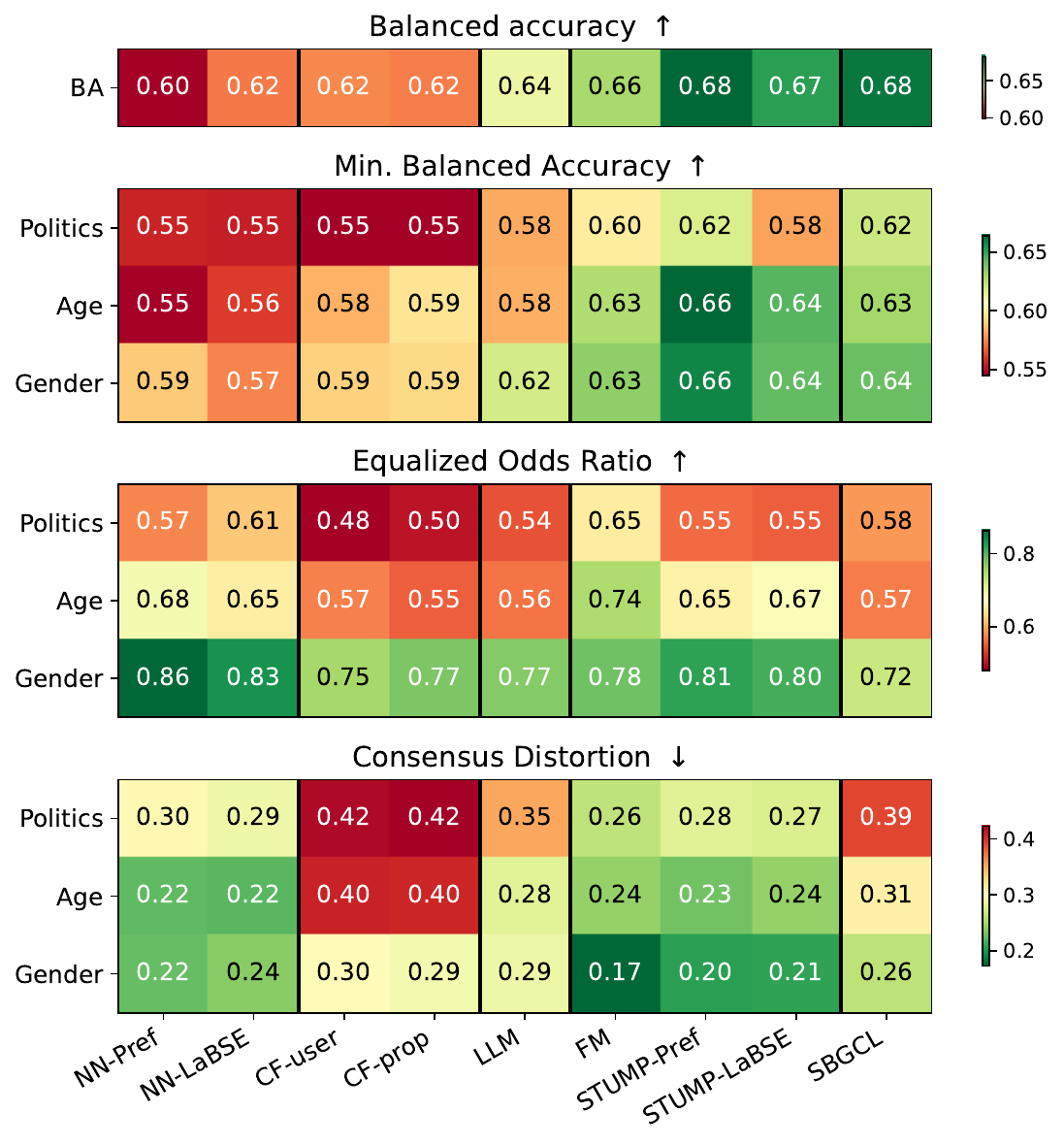}
    \caption{Equalized Odds Ratio (\textuparrow), Min. 
    B-Acc (\textuparrow) and CD (\textdownarrow) averaged over campus-protests, right-to-assemble, and steuer-debate. B-Acc is shown for reference.}
    \label{fig:distortion_3_metrics}
\end{figure}

% \paragraph{Does higher balanced accuracy lead to a less distorted landscape of preferences ?}
\paragraph{Accuracy vs. Opinion Distortion}
% Figure \ref{fig:mean_scatter} displays the Balanced Accuracy compared to distortion and fairness metrics per model. Though SBGCL performs the best regarding BA (getting the best results on 10 of the 14 datasets, and the best average), it performs 5th out of 8 in ARHR and CS and 6th on CD (politics). 
% More generally, we find that there is little to no correlation between BA and our metrics ($R^2=0.05$ for ARHR, $R^2=0.14$ for CS, $R^2=0.13$ for CD on the political attribute. Results for all metrics are given in the Complementary Materials). We only find a strong correlation between BA and min-BA, which is to be expected.  
% \newline
% Figure \ref{fig:distortion_3_metrics} provide the averaged results on the attribute-based metrics (EOR, Min BA and CD) per model, for the three attributes shared by the datasets. This complete these findings, with SBGCL not performing as good as other models, despite showing the best BA. 
Figure \ref{fig:mean_scatter} presents B-Acc compared to distortion and fairness metrics per model. Although SBGCL achieves the highest B-Acc overall, it falls short on our proposed metrics, ranking 5th on ARHR and CS and 6th on CD. More generally, we find little to no correlation between B-Acc and these metrics ($R^2=0.05$ for ARHR, $R^2=0.14$ for CS, $R^2=0.13$ for CD on the political attribute)\footnote{We provide spearman correlation between all metrics used in this work in the supplementary materials.}, with the exception of Min. B-Acc, which is naturally strongly correlated. Figure \ref{fig:distortion_3_metrics} displays model performance on the attribute-based metrics (EOR, Min. B-Acc and CD) across the three common attributes (Gender, Age, Politics), further supporting the finding that SBGCL lags behind other models despite its high B-Acc.

% \paragraph{Do existing PI methods provide acceptable results regarding opinion distortion ?} 
\paragraph{Are Existing PI methods Sufficient?}
% or \paragraph{Performance of Existing PI methods}
% No model perform noticeably better than the others. NN models perform well regarding CD and ARHR, but have limited balanced accuracy results and perform the worst on CS. both CF models show high BA scores, especially given that they are not learned, but obtain the worst results in almost all distortion and fairness metrics. Overall, STUMP-based models seem to provide the best trade-off: they show results slightly below SBGCL, but distort the opinion landscape significantly less. 
% \newline
% In general, results on opinion landscape distortion are not acceptable. No model provides statistically plausible approval rates: no model, on no dataset, obtains an ARHR score above 0.95, which would be the highest expected score given that we work with a 95\% confidence interval. Similarly, the lowest average CD is 21.7 (obtained by FM). Such a difference in average approval rate is unacceptable for democratic processes. Finally, CS is overall low, the best performing model (CF) obtaining a 22.9 average. 
No model noticeably outperforms the others across the board. NN models achieve strong CD and ARHR results, but exhibit limited B-Acc and score the lowest on CS. Both collaborative filtering models show high B-Acc. scores—particularly impressive given that they are not learned—yet obtain the poorest results on almost all distortion and fairness metrics. Overall, STUMP-based models provide the best trade-off, showing accuracy results slightly below SBGCL, but at the same time distorting the preference landscape significantly less. 
In general, current results regarding preference landscape distortion fall short of acceptable standards for democratic processes. No model produces statistically plausible approval rates (at a 95\% confidence interval); specifically, no model achieves an ARHR score above 0.95 on any dataset (cf. Supplementary Material). Similarly, the lowest average CD is 21.7 (obtained by FM) which is arguably too high for democratic processes. Finally, CS remains consistently low, with the best performing model (CF) achieving an average of only 22.9. Additionally, a per-language analysis on \emph{Eurhope} (reported in the Supplementary Material) shows disparities in treatment. Ultimately, these findings reveal a critical limitation in current preference inference methods, underscoring the need for improved approaches that prioritize faithful representation and preservation of the underlying preference landscape alongside predictive accuracy.

\begin{figure}[t]
    \centering
    \includegraphics[width=.99\linewidth]{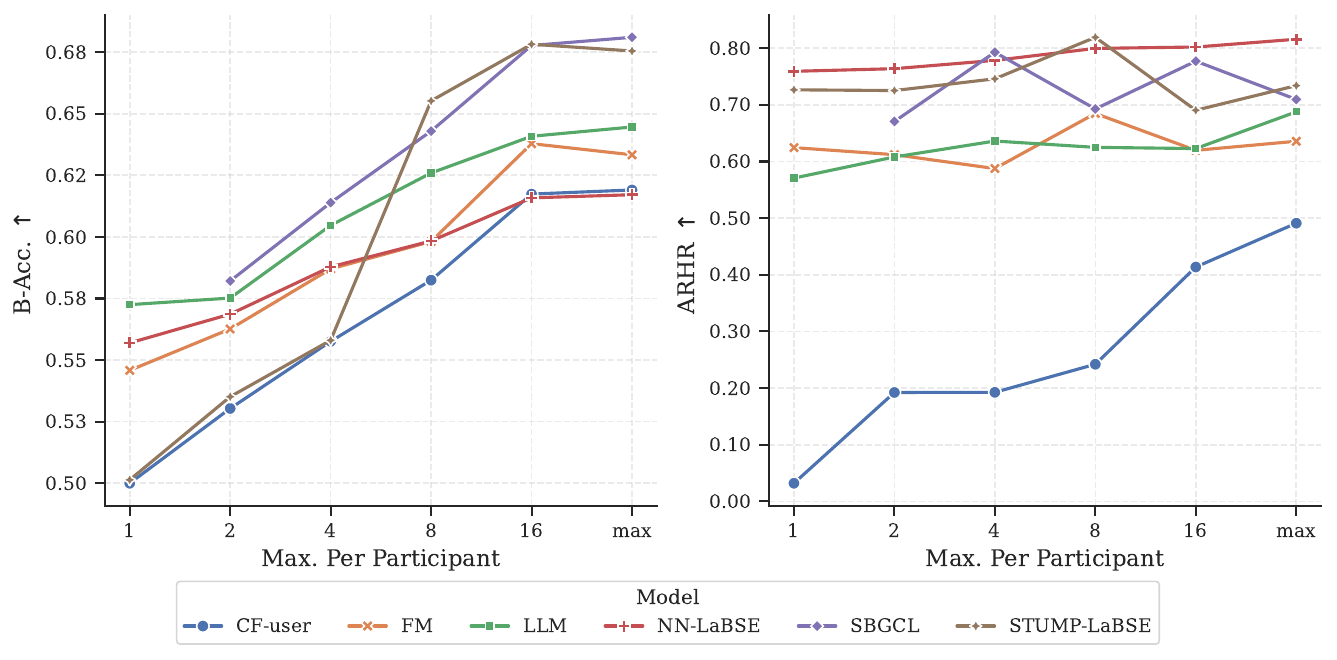}
    \caption{
    % B-Acc and ARHR averaged over three datasets, when limiting the number of examples per user to different values.
    B-Acc. and ARHR averaged over three datasets across varying numbers of examples per user.}
    % \itodo{en moyenne sur tout sauf eurhope}
    \label{fig:subsample}
\end{figure}

\paragraph{Data Efficiency} Following \citet{konya-etal-2022-elicitation}, we evaluate the effect of preference sparsity by limiting the number of observed preferences per participant (1, 2, 4, 8, 16, or all available)\footnote{SBGCL requires at least two observed preferences per participant and is therefore not evaluated with one.} on the three datasets with socio-demographic attributes. Figure~\ref{fig:subsample} reports B-Acc and ARHR (results for the other metrics are provided in the Supplementary Material).
%B-Acc, Min. B-Acc, and CD consistently improve as more preferences become available, contrasting with \citet{konya-etal-2022-elicitation}, who report stable STUMP performance from only five preferences per participant.\footnote{Their evaluation relies on standard accuracy, whereas we use B-Acc, which is better suited to these imbalanced datasets.} 
B-Acc, Min. B-Acc, and CD consistently improve as more preferences become available, an important result not observed when evaluating using only accuracy (\citet{konya-etal-2022-elicitation}).\footnote{\citet{konya-etal-2022-elicitation} report no improvement in accuracy above five preferences per participant, but do not evaluate using collective-centric metrics.} This further highlights the need for collective-centric metrics beyond accuracy. In contrast, ARHR, EOR, and CS remain largely unchanged, with the latter exhibiting noisy but uniformly low values (0--0.18). These findings suggest that several limitations of current preference inference models are structural rather than a consequence of sparsity.

\section{Conclusion}
Participatory democracy platforms enable collective decision-making among thousands of participants, yet the resulting voting data are often highly sparse. Preference inference is therefore essential for reconstructing the preference landscape and supporting downstream decision-making.

We introduce a collective-centric evaluation framework, grounded in deliberative democratic theory, to assess whether preference inference systems faithfully preserve the collective structure of expressed preferences rather than merely predicting individual votes accurately. Our findings demonstrate that preference inference is not a neutral computational step: models with high predictive accuracy can substantially distort patterns of consensus and the preferences of particular groups. Concerningly, across all evaluated consultations, no existing method faithfully preserves the underlying preference landscape. Alongside this evaluation framework, we release a new multilingual benchmark spanning four consultations, nearly 90,000 participants, one million votes, and 22 languages. In particular, the Eurhope consultation constitutes the largest publicly available consultation dataset to date, comprising eight times more participants and twice as many propositions as the largest previously available benchmark, while being the first multilingual.

This work establishes collective-centric evaluation as a standard objective for future preference inference systems, encouraging models that optimize not only predictive accuracy but also faithful collective representation.

\subsection{Limitations and Ethical Considerations}
Although our framework provides principles for evaluating preference inference systems, it should not be interpreted as legitimizing preference inference as a substitute for genuine democratic participation. Preference inference can only complement citizen consultations whose design already ensures broad, representative, and meaningful participation. Moreover, our evaluation framework operationalizes a specific set of democratic principles grounded in deliberative theory, but these principles are not exhaustive. Alternative conceptions of deliberation and democratic representation may motivate additional evaluation criteria.

\subsection{Acknowledgments}
This research was funded by BPI-France under the project AI For Democracy - Democratic Commons, one of seven winners of BPI-France's 'Digital Commons for Generative AI' call for projects, conducted as part of the France 2030 investment plan. We would like to thank the organizations that generously shared data for this research. We are grateful to Netzwerk Steuergerechtigkeit and Mehr Demokratie for providing data from the German \textit{Steuer-Debatte} project, and to the Secrétariat général de la défense nationale (SGDSN) for providing data from the French citizen consultation on foreign interference (\textit{Lutter contre les ingérences numériques}). Their contributions made it possible to evaluate our approach on real-world citizen participation datasets. We also thank the members of the \textit{Democratic Commons} project, especially Manon Berriche, for their help and careful proof-reading. This work was performed using HPC resources from GENCI–IDRIS (Grant 2026-A0201017543). 

\bibliography{aaai2027}


\end{document}